\documentclass[aps,prd,twocolumn,nofootinbib,groupaddress,longbibliography]{revtex4-1}

\usepackage[T1]{fontenc} 
\usepackage{amsmath,amsfonts,amssymb,graphicx,color,times,bbm,psfrag,dsfont}
\usepackage{float}
\usepackage[caption=false]{subfig}
\usepackage{siunitx}
\definecolor{myblue}{rgb}{0.1,0.24,0.6}
\definecolor{myred}{rgb}{0.6,0.1,0.2}
\usepackage[colorlinks=true, citecolor=myblue, linkcolor= myblue, urlcolor=myblue,bookmarks=true]{hyperref}

\def \beq{\begin{equation}}
\def \eeq{\end{equation}}
\def \bea{\begin{eqnarray}}
\def \eea{\end{eqnarray}}
\def \bem{\begin{pmatrix}}
\def \eem{\end{pmatrix}}
\def \bs{\boldsymbol}
\def \nn{\nonumber}

\def \magic{\theta_{\text{\,magic}}^{\text{\,eff}}}

\newcommand{\ket}[1]{\left \rvert #1 \right \rangle}
\newcommand{\bra}[1]{\left \langle #1 \right \rvert}

\begin{document}

\title{Pressure-Induced Metal-Insulator Transition in \\ Twisted Bilayer Graphene}
\author{Bikash Padhi}
\thanks{\href{phys.bikash@gmail.com}{bpadhi2@illinois.edu}}
\affiliation{Department of Physics and Institute for Condensed Matter Theory, University of Illinois at Urbana-Champaign, 1110 W. Green Street, Urbana, IL 61801, USA.}
\author{Philip W. Phillips}
\thanks{\href{dimer@illinois.edu}{dimer@illinois.edu}}
\affiliation{Department of Physics and Institute for Condensed Matter Theory, University of Illinois at Urbana-Champaign, 1110 W. Green Street, Urbana, IL 61801, USA.}
%
%
%
%

\begin{abstract}
Recent experiments on twisted bilayer graphene (TBLG) have observed insulating states for two and three unit charges per moir\'e supercell, whereas the quarter--filling state (QFS) remained metallic. Subsequent experiments show that under hydrostatic pressure the QFS turns insulating for a certain window of pressure. In fact, the resistivity of the 1/2--filling and 3/4--filling states are also enhanced in the same pressure-window. Using pressure-dependent  band structure calculations we compute the ratio of the potential to the kinetic energy, $r_s$.  We find a window of pressure for which $r_s$ crosses the threshold for a triangular Wigner crystal, thereby corroborating our previous work that the insulating states in TBLG are driven by Wigner physics,   A key prediction of this work is that the window for the onset of the hierarchy of Wigner states that obtains at commensurate fillings conforms to a dome shape under pressure. We also predict the optimal condition for Wigner crystallization to be around $1.5 $ GPa. Consequently, TBLG provides a new platform for the exploration of Wigner physics and its relationship with superconductivity. 
\vspace{4mm} \\
\end{abstract}

\maketitle

\section {Introduction}

Twisted bi-layer graphene (TBLG) is a true example of emergence.  Electrons in single layers of
graphene are free while those in the composite consisting of two layers twisted close to the magic
angle such that the electronic bands are essentially flat have almost no kinetic energy, $E_K$. In
such cases, the physics is dominated by the interactions, $E_U$ , between the electrons.  The experimental observation of correlated insulating phases and superconductivity is hence not unexpected.  As a result of these discoveries~\cite{cao2018Mott,cao2018SC}, TBLG is largely viewed as a problem in strongly correlated physics~\cite{ourNano, LeeKekule, SpinLiq, LiangFuCDW, FanYang, Ochi, Subir, Kivelson,balents,AshvinPo,LiangMIT, baskar, KoshinoFu, AsvinSenthil, Oskar, FaithfulAshvin, Fernandes}. However, unlike conventional strongly correlated materials such as the cuprates or the heavy fermions, TBLG offers an extremely tunable platform. Namely, through the twist angle one can control the extent of strong correlation. 

When two layers of graphene are rotated with respect to each other, a so called moir\'e lattice emerges~\cite{LopesPRB,Mele10,LopesPRL,MCDMoire}, which is a triangular lattice with periodicity $\lambda_s=a/(2\sin\theta/2)$. Here $a = \SI{2.46}{\angstrom} $ is the lattice constant of pristine graphene layers. This emergent lattice has an approximate SU(4) symmetry due to the valley and spin degeneracies. Thus a moir\'e band can hold up to four electrons. So if we consider a moir\'e supercell of area $A_s=\sqrt{3}\lambda_s^2/2$, the superlattice density ($n_s$) can be fixed using $A_s n_s=4$.  Consequently, it is convenient to define the index $\nu=n_e A_s$ which serves as the electron filling factor.  The initial experiments~\cite{cao2018Mott,cao2018SC} in this regime showed that insulating states can arise for $\nu=\pm 2,3$.  Doping away from $\nu = -2$ resulted in superconductivity with a transition temperature of $1.5\,$K. These results were later confirmed by various groups~\cite{YoungDean, PasupathySTM, Gil, Kastner, EfetovSC, gordon}. In particular, it was demonstrated~ \cite{YoungDean} that hydrostatic pressure can also be used to further tune the effects of twist angle. This gave way to certain metal--insulator and insulator--metal transitions which were not observed at ambient pressure. In this work we try to address the mechanism behind these peculiar transitions.

At zero temperature, a measure of the degree of correlation can be, $E_U/E_K \equiv r_s$. Starting from a two-dimensional homogenous gas of electrons (2DEG) one can drive the system through different phases simply by tuning $r_s$. This is true because the energy of a many-body ground state, $E_{0}(r_s)$, is solely a function of $r_s$. The two asymptotic phases one thus obtains are a Fermi liquid phase for $r_s \lesssim 1$ and a Wigner solid phase~\cite{Wigner34} for $r_s \gtrsim 37$~\cite{David37}. Experimentally, in principle, one can access these phases by changing the carrier density ($n_e$) or applying a magnetic field ($B$). However, in the case of TBLG, $r_s$ can also be tuned by the twist angle ($\theta$) or hydrostatic pressure ($P$).  At ambient pressure, a single layer of graphene~\cite{NoWigner} or a Bernal stacked bilayer of graphene~\cite{E-field, DahalInhomo} has $r_s \lesssim 1$. Twisting the layers towards a magic angle configuration increases $r_s$ of this TBLG system, driving it towards a Wigner phase~\cite{ourNano}. This can simply be understood by the flattening of the moir\'e bands. In fact, the proclivity of flat-band systems to form Wigner crystals has not gone unnoticed~\cite{CongjunWu1}. However, a natural question that arises is, how does pressure modulate $r_s$? To answer this question, we numerically compute $r_s$ of TBLG as a function of $\theta, n_e, P$. With this we demonstrate that the metal--insulator--metal transition mentioned above can be understood as a melting of a Wigner solid phase; see Fig.~\ref{fig:rsP}.
 
This paper is organized in the following manner. In Sec. \ref{sec:MottWC} we first argue that the correlated insulators observed in TBLG are Wigner, not Mott insulators. In Sec. \ref{sec:Model} we discuss the tight-binding Hamiltonian we use for computing the band structure of TBLG. Pressure is then incorporated into computing the band structure in the presence of triangular warping, and a pressure-dependent effective magic angle is obtained in Sec. \ref{sec:Magic}. We then proceed to compute pressure dependence of $r_s$ at various commensurate fillings in Sec. \ref{sec:rs}. In Sec. \ref{sec:Correction} we discuss a few possible corrections to our estimation of $r_s$. We conclude our discussion in Sec. \ref{sec:Discuss} by commenting on a few other aspects of TBLG in relation to Wigner crystals (WCs). 

\section{TBLG: Mott versus Wigner Paradigm}
\label{sec:MottWC}

This paper addresses the pressure dependence of the insulating states.  Electronic band structure calculations~\cite{PabloPress, FangKaxiras, KoreanPress} as a function of pressure in TBLG offer immediate insight into the physics at play. Hydrostatic pressure causes uniaxial compression between the graphene layers~\cite{hBNphase}, which in turn increases the interlayer tunneling, thereby changing the magic angle condition.  However, an additional feature also appears: the bandwidth shows a dome-like shape with increasing pressure.  Because the interactions remain fixed, the ratio $r_s$ increases, favoring Wigner crystallization. Although, simply from the bandwidth perspective, Mott insulation might also seem favorable.  

It is important then to determine what physics TBLG exhibits that conforms to either scenario. Cao, \textit{et al.} as well as others~\cite{balents,AshvinPo,LiangMIT,baskar} attributed the insulating states at $\nu=\pm 2,3$ to Mott physics.  Within this paradigm, insulating behavior should exist whenever the band is partially filled.  However, metallic not insulating behavior exists at $\nu=1$ in the experiments of~\cite{cao2018Mott,cao2018SC}.   This is a potential problem for the application of the Mott scenario to TBLG.    It is not surprising then that the Mott criterion~\cite{MottDavis} $n_e^{1/d}a_0^\ast\approx O(1)$ is not satisfied in TBLG near the magic angle.  In fact, this is satisfied only for $\theta-\theta_{\rm magic} \gtrsim 0.7$.  

Another key distinguishing feature between a Mott and a pinned Wigner insulator is that the spatial symmetry of the Mott state is always the same as that of the lattice. However, a Wigner crystal, being an emergent lattice by itself, may or may not adhere to the symmetries of the underlying lattice. Within the Mott paradigm, the relevant question is how can the electrons be placed in a moir\'e lattice without creating a new electron lattice distinct from the underlying triangular moir\'e lattice.  That is, because of the Coulomb interaction, the electrons must occupy spatially separated locations in each moir\'e cell regardless of the underlying SU(4) symmetry.  Consequently, except, for $\nu=1$, any arrangement of the electrons must create a lattice distinct from the triangular lattice. The honeycomb ($\nu=2$) (also proposed previously as a possible ground state~\cite{Subir,LeeKekule}) and kagome ($\nu=3$) lattices are examples of Wigner lattices as they all break the underlying triangular symmetry~\cite{ourNano}. While the most common instances of Wigner crystal formation involve a magnetic field~\cite{MonarkhaRev} that quenches the kinetic energy, the situation in TBLG is not very dis-similar because it is well known that a relative twist between two layers of graphene generates~\cite{NonAbelianPRL} a non-Abelian gauge pseudo-potential~\cite{NABexp} with a magnetic length equal to the moir\'e lattice constant.  

In the following sections we demonstrate that increasing the pressure in TBLG leads to $r_s>37$, thereby resolving the pressure-induced metal-insulator transition in TBLG. We map out the phase diagram using realistic parameters for TBLG and determine the regime where the $\nu=1$ state crosses the WC threshold.  We find that the quarter-filling state in the new experiments~\cite{YoungDean} at $1.33$ and $2.21\,$GPa are well within the Wigner regime while those at ambient pressure correspond to $r_s < 37$. We also confirm the experimental trend that hydrostatic pressure enhances the insulating states at $\nu=2,3$.

\section{The Tight--Binding Hamiltonian}
\label{sec:Model}

We start by computing the pressure-dependent band structure and subsequently $r_s$. In our discussion, we will focus explicitly on device D2 of~\cite{YoungDean}. Consider two layers of graphene, each rotated by $\pm \theta/2$ around an axis passing through an A$_1$B$_2$ site, where the subscripts denote the layers, and A, B are sublattice labels. When $\theta$ is small, the supercell consists of a large number of atoms, $\sim 10^4$, making \textit{ab initio} methods~\cite{ShallcrossPRL} less viable or reliable~\cite{overlapNano} than the tight-binding schemes~\cite{LopesPRL, LopesPRB}. Here, we follow the tight-binding scheme of~\cite{PabloPress}, where the tight-binding parameters are functions of pressure.  Also, since we work with a tight-binding model, unlike the case of a continuum model, we limit our discussion to commensurate structures obtained for twist angles~\cite{Shallcross10},
\beq
\theta = \cos^{-1} \left[ \frac{m^2 + 4mn + n^2}{2(m^2 + mn + n^2)} \right] \, , \quad m, n \in \mathrm{Z} \, .
\label{commensurate} 
\eeq
The twist angle of the D2 sample is $\theta=\ang{1.27}$, which is not a commensurate angle. Because of the reasoning above, we work with the nearest commensurate angle, $\theta \approx \ang{1.25}$, obtained for $(m, n)=(26, 27)$.

We denote the supercell vectors as $\bs R_1 = m \bs a_1 + n \bs a_2 $ and $\bs R_2 = - n \bs a_1 + (m+n) \bs a_2$, with $\bs a_1$, $\bs a_2$ being the lattice vectors of original graphene layer, and each unit cell is $|m-n|$ ($=1$ for D2) times larger than the moir\'e periodicity, $\lambda_s$. For commensurate structures, there is a well defined moir\'e Brillouin zone (MBZ). The symmetry points of the MBZ will be labeled as $\bar{\Gamma}$ (zone center), $\rm \bar{M}$ (edge center), and $\rm \bar{K}$ (zone corner). Since tunneling between two valleys is prevented in a low-energy description ($\lesssim 1\;$eV) and as a result of the valley degeneracy, our calculation only considers an MBZ formed near the $\bs K$ (Dirac) point of the original lattice.

We begin with a simplified description, ignoring any angular dependence of the hybridization or the orbital overlaps. The generic non-interacting part of the Hamiltonian is 
\beq
H = - \sum_{i, j} t(\bs R_i - \bs R_j) \ket{\bs R_i} \bra{\bs R_j} + \text{H.c.} \, ,
\label{TBHamil}
\eeq
where $\bs R_i = \sum_{x,y,z} {R_i}^a \, \bs e_a$ is the atomic coordinate in the basis of $\{\bs e_a\}$, and $ \ket{\bs R_i}$ is the wave function at site $i$. The tunneling strength between sites $i$ and $j$ is measured by the tight-binding parameter $t(\bs R_i - \bs R_j)$. We can express this tight-binding parameter using a simple linear combination of $p_z$ orbitals as 
\begin{subequations}
\begin{gather}
t(\bs R) = V_{pp\pi} (\bs R) \, \sin^2 \gamma + V_{pp\sigma} (\bs R) \, \cos^2 \gamma \, ,  \\ 
R \cos \gamma = \bs R \cdot \bs e_z  \, .
\label{eq:tperp}
\end{gather}
\end{subequations}
Here $R$ is the length of the vector $\bs R$ joining two atoms and $\bs e_z$ is the unit vector along the $c$ axis. The overlap or transfer integrals, $V(\bs R)$, can be expressed in terms of the Slater-Koster parameters~\cite{SlaterKoster}
\bea
V_{pp\pi} (\bs R) &=& - t_0 \exp\left( - \frac{R - a_0}{r_0} \right) \, , \\ 
V_{pp\sigma} (\bs R)  &=& t_\perp \exp\left( - \frac{R - d_\perp}{r_0} \right) \, .
\label{SlaterKoster}
\eea
$r_0 = 0.318 \, a_0$ is an isotropic decay length chosen~\cite{overlapNano} for the transfer integrals so that the next-nearest in-plane overlap becomes $0.1t_0$~\cite{nnVpp}. $t_\perp$ is the $\sigma$--bond (or interlayer coupling) strength between the $sp_2$ orbitals of the AB stacked bilayers. At ambient pressure $t_\perp|_{P = 0} \equiv t^0_\perp \approx 0.31\, \rm eV$. $t_0 \approx 2.7\;$eV is the in-plane $\pi$--bond strength of the two neighboring $p_z$ orbitals (separated by $a_0 = a/\sqrt{3} = \SI{1.42}{\angstrom}$) in  single-layer graphene. We use $d_\perp$ for the inter-layer spacing at finite pressure $P \, $(GPa) and $d_\perp^{\,0} = \SI{3.35}{\angstrom}$ is the spacing at ambient pressure.

\begin{figure}[t!]
\centering
\includegraphics[width=0.9 \columnwidth]{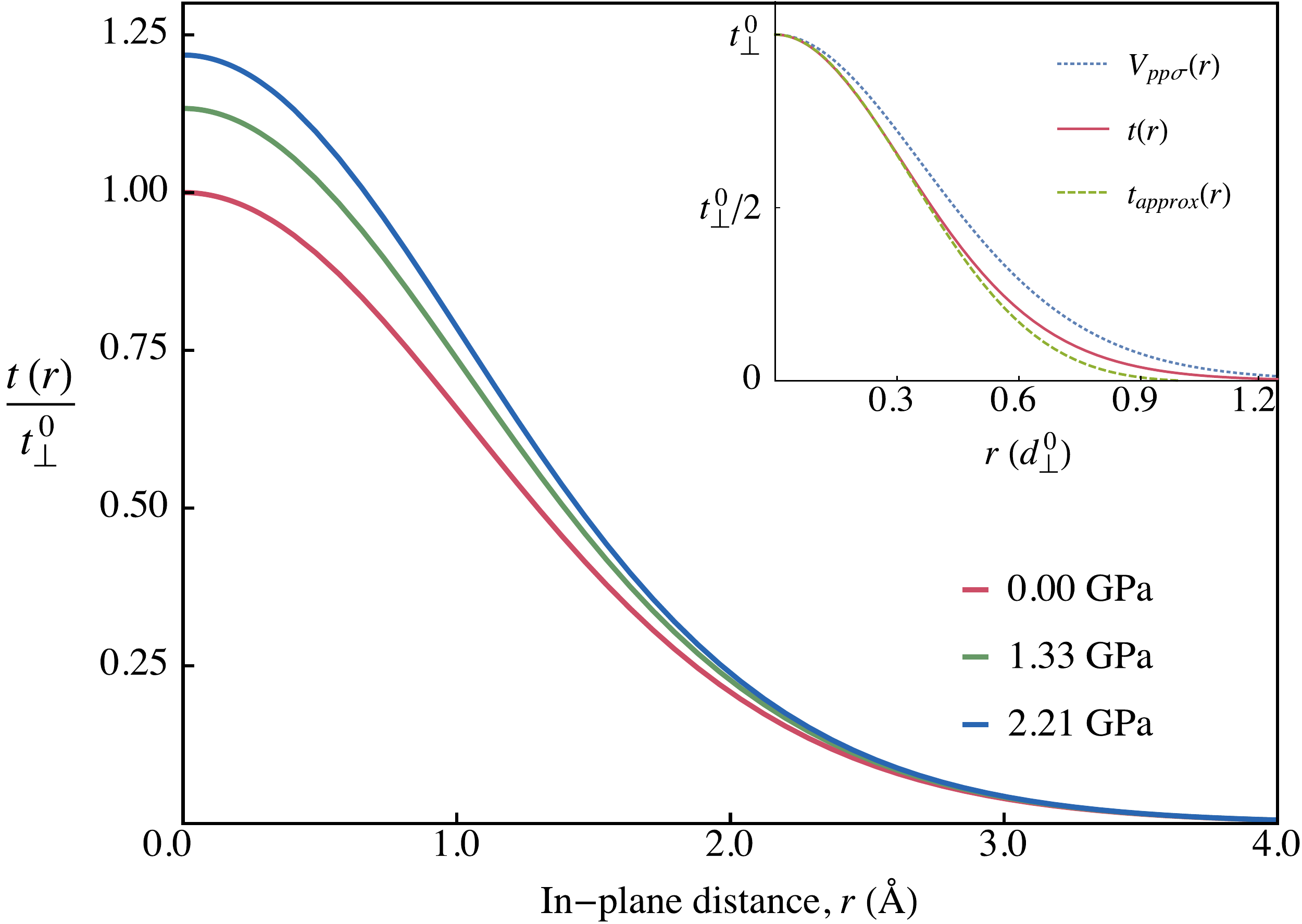} 
\caption{Inter-layer tunneling for different pressures as a function of distance from the site of rotation. With increasing pressure, the inter-layer distance decreases causing the tunneling strength to increase. Inset: The dominant contribution to $t(\bs r)$ comes from $V_{pp\sigma}$ which leads to a simpler expression in Eq.~\eqref{approxt(R)}.}
\label{t(R)press} 
\end{figure}

Since $d_\perp/ a_0 \gtrsim 2 $, near the stacking center, the tunneling parameter $t(\bs R)$ is largely dominated by the $\sigma$ bond, and thus, the function $t(\bs R)$ can be approximated as 
\beq
t(\bs r) \approx t_\perp \left( 1 - \frac{r^2}{d_\perp^2}   \right) \exp\left(-\frac{r^2}{2r_0 d_\perp }\right) \, .
\label{approxt(R)}
\eeq
In the inset of Fig.~\ref{t(R)press}, the behavior of Eq.~\eqref{approxt(R)} is juxtaposed with the exact result from Eq.~\eqref{eq:tperp}, which shows an exponential reduction of the tunneling strength for $r \gtrsim d_\perp$. This also causes the Fourier transform to sharply decay for any $k \gtrsim 1/d_\perp$. Thus, for a low-energy model, it is sufficient to work with $t_\perp(\bs K)$ only and not include the higher modes, such as $t_\perp(\bs K + \bs G)$, where $\bs G$ is a moir\'e reciprocal lattice vector. One can perform a Fourier transform of $t(\bs r)$ computed above to determine $t_\perp(\bs K)$, or since we work in the $\theta \sim 1^\circ$ limit, (for AB stacking) one can approximate $t_\perp(\bs K)/ A_0 \, \equiv w = \frac{1}{3}t_\perp \, .$ Here $A_0 = \sqrt{3} a^2/2$ is the area of the single-layer graphene unit cell and the factor of $3$ takes into account that there are three equivalent Dirac cones. One can use $w$ as the input parameter in the effective theories.

In concluding this section we note that, in our discussion, $t_\perp$ is the single energy parameter that is affected by pressure [see Eq.~\eqref{effective1para}]. The in-plane energetics, controlled by in-plane hopping, may change under very high pressure, especially in the presence of a hexagonal boron nitride (hBN) substrate; however, for the range of pressure relevant here, such effects can be safely neglected~\cite{hBNphase}. 

\section{Pressure Dependence of Magic Angle}
\label{sec:Magic}

In order to quantify the effect of pressure on $t_\perp$ first we need to obtain the relation between $d_\perp$ and pressure. Application of hydrostatic pressure can readily reduce $d_\perp$~\cite{hBNphase}, the experimental consequences of which have been studied in~\cite{YoungDean}. This compression factor, denoted by $\delta_d$, is related to applied pressure through the Murnaghan equation of state~\cite{KoreanPress}
\begin{align}
1 - \frac{d_\perp}{d_\perp^{\, 0}} \equiv \delta_d  = 10.48 \ln \left( 1 + \frac{P}{5.73} \right) \% \,.
\label{Murnaghan}
\end{align}
The numbers appearing here are fixed using density functional theory~\cite{PabloPress}. An immediate consequence of a reduced $d_\perp$ is an enhanced magic angle, which we denote by $\magic$.  In fact, for experimentally accessible pressures, this mechanism can enhance $\magic$ up to $\ang{3}$. The primary advantage of a large $\magic$ is an enhanced Coulomb energy scale ($E_U \sim \lambda_\theta^{-1} \sim \theta$) which could also result in an increased $T_c$~\cite{YoungDean}.

\begin{figure}[t]
\centering
\subfloat{\includegraphics[width=0.5 \columnwidth,height=0.42
\columnwidth]{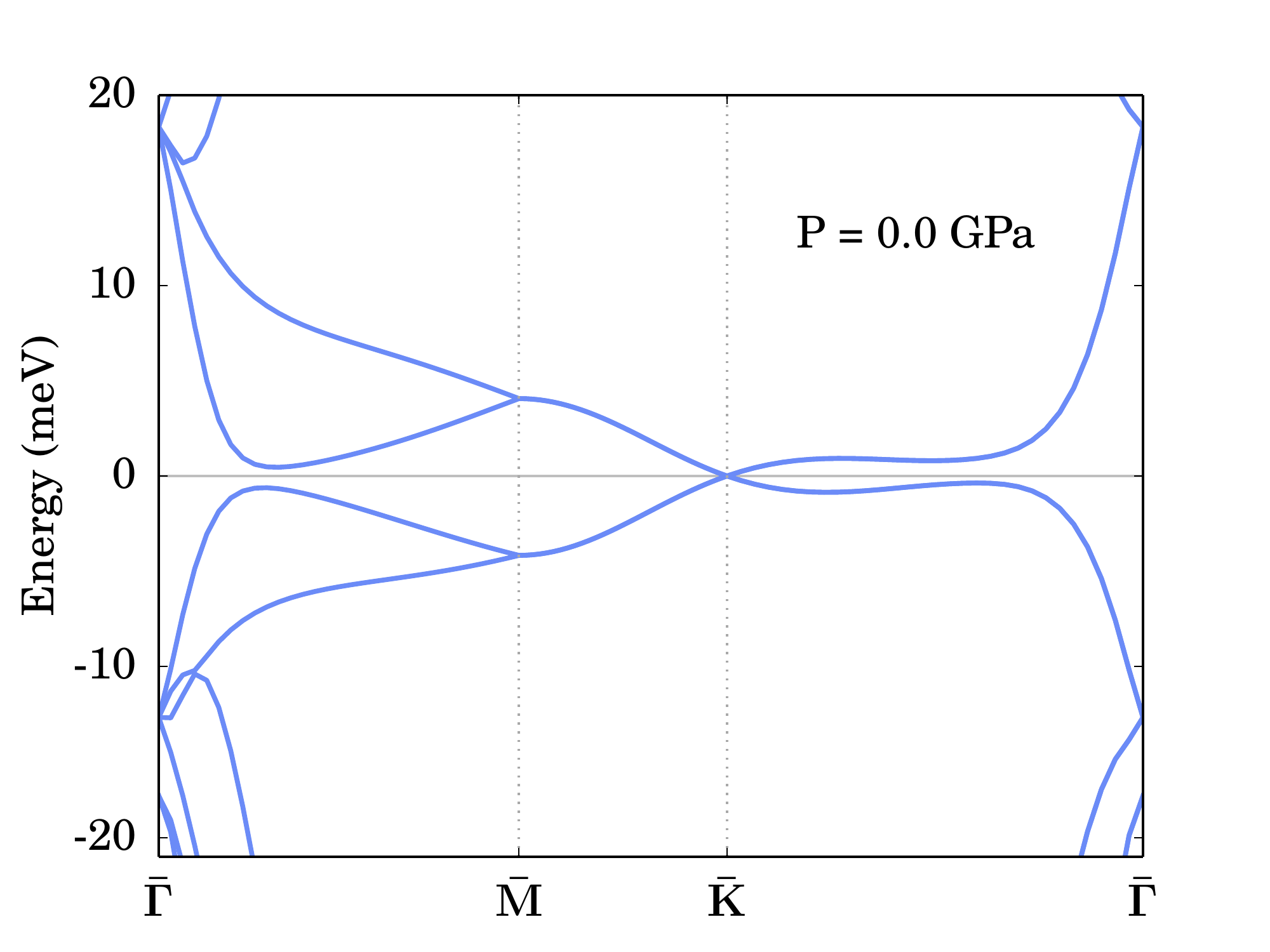} }
\subfloat{\includegraphics[width=0.5 \columnwidth,height=0.42
\columnwidth]{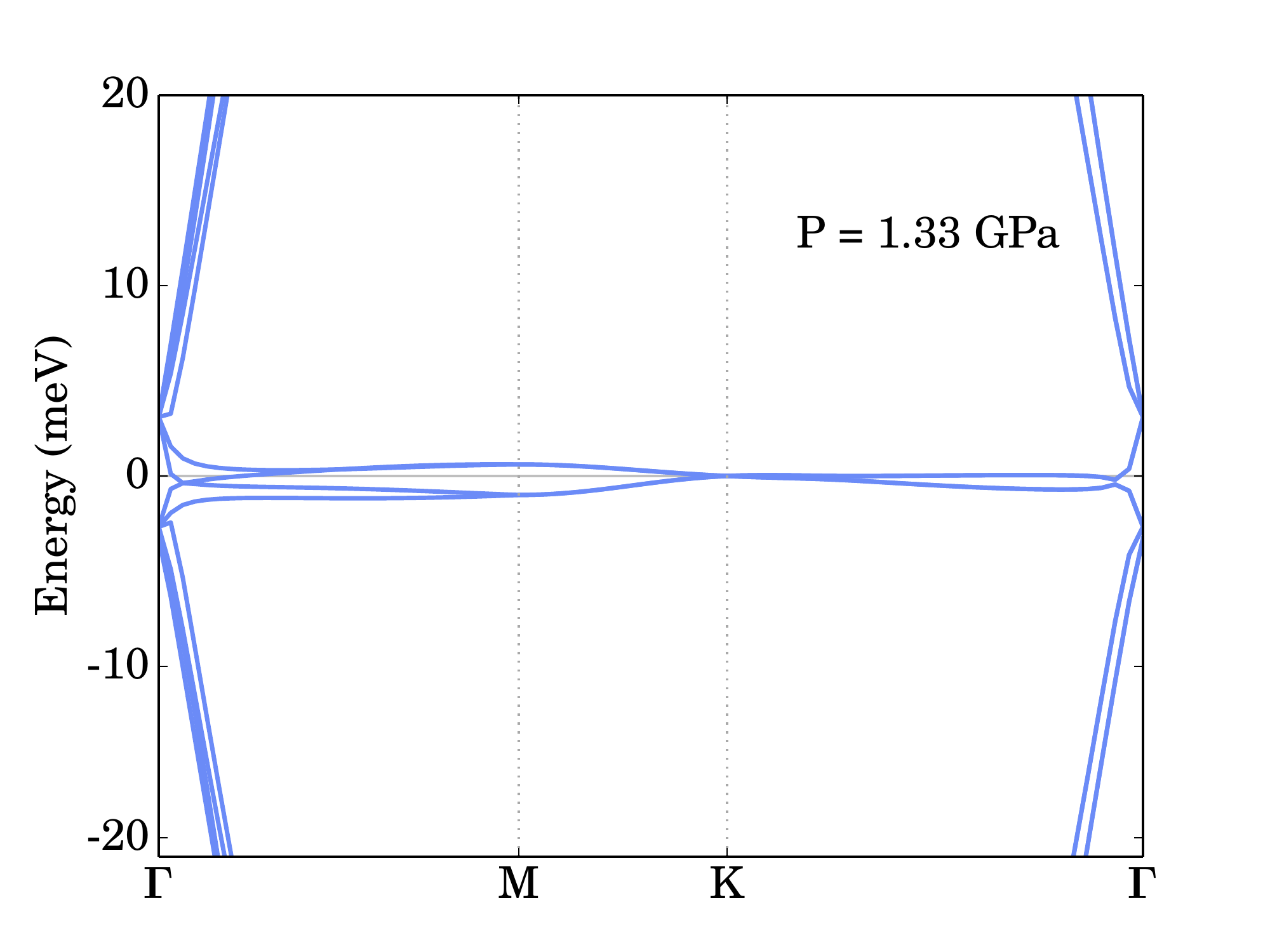} } \\
\subfloat{\includegraphics[width=0.5 \columnwidth,height=0.42
\columnwidth]{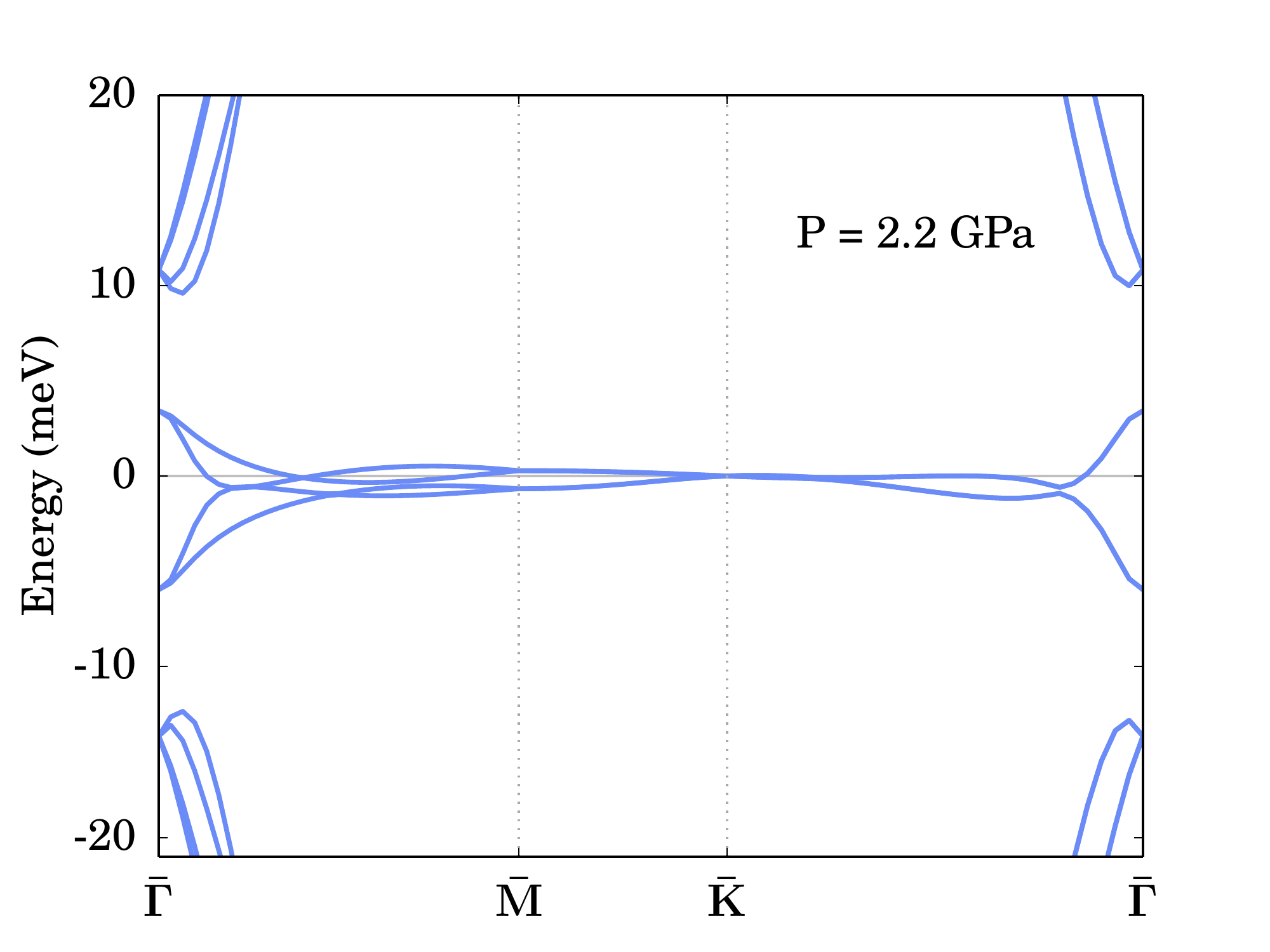} } 
\subfloat{\includegraphics[width=0.5 \columnwidth,height=0.42
\columnwidth]{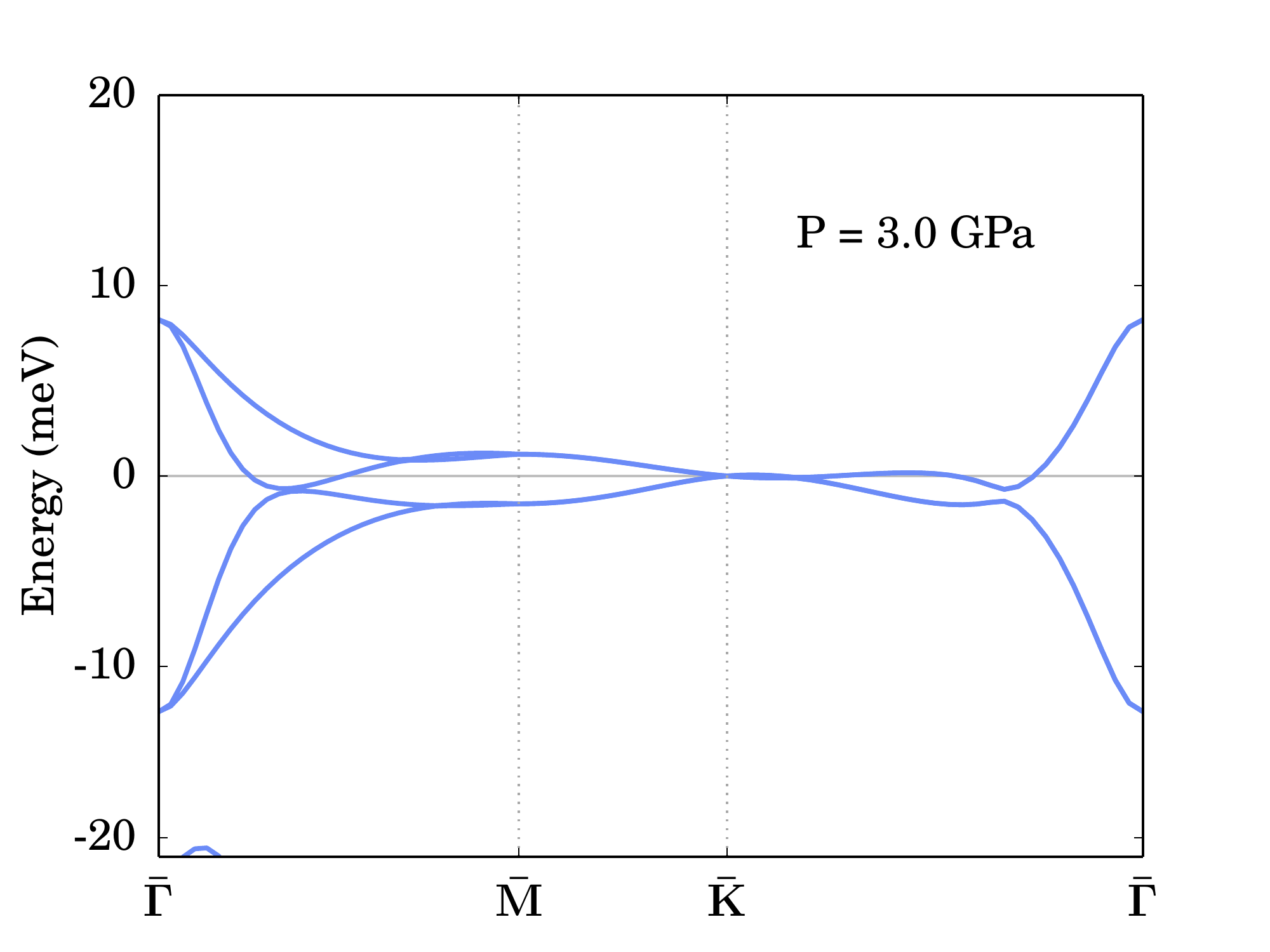} }
\caption{
Pressure-dependent band dispersion of TBLG for $\theta = \ang{1.25}$. The parameters used in obtaining these are listed in the Table~\ref{tablepara}. With increasing pressure, the low energy bands become flatter; however, beyond $\sim 1.45 \pm 0.1$ GPa, the bandwidth  increases subsequently. The reason behind such an optimal behavior can be understood from Eq.~\eqref{effectivemagic}.}
\label{figBS} 
\end{figure}

In order to express $t(\bs R)$ as a function of $\delta_d$ (hence, $P$), we use Eq.~\eqref{Murnaghan} in Eq.~\eqref{SlaterKoster} or \eqref{approxt(R)} and rewrite the tight-binding Hamiltonian. However, at finite pressure the Slater-Koster approximation turns contentious as the overlap between the Wannier orbitals develops a strong angular dependence~\cite{PabloPress, FangKaxiras, KoreanPress}. This is a result of the overlap of Wannier orbitals with angular momentum, $m = 3 n$, with $n \in \mathbb{Z}$, where $3$ appears due to the $D_3$ point group symmetry of the underlying lattice. We will denote the radial components of such overlap functions with $V_{|m|} (r)$, whereas the angular dependence simply will be $\cos(m\theta)$. In this notation, Eq.~\eqref{eq:tperp} can be viewed simply as $V_0(r)$, which still is the leading contribution to $t_\perp(\bs r)$. In fact, we will only consider the overlaps from the $m = \pm 3, \pm 6$ orbitals since the effects from the overlap of the higher-order orbitals are negligible~\cite{FangKaxiras}. A  real space expansion of $t_\perp(\bs r)$ is thus written as~\cite{PabloPress, FangKaxiras, KoreanPress, FangKaxiras}
\bea
t_\perp(\bs r) = V_0 (r) + V_3(r) \left[ \cos(3\theta_{12}) + \cos(3\theta_{21}) \right] \nn  \\ + V_6(r) \left[ \cos(6\theta_{12} + \cos(6\theta_{21})) \right] \, ,
\label{angularTB}
\eea
where $\theta_{ij}$ are the angles between the vectors connecting the $i\textsuperscript{th}$ site to the $j\textsuperscript{th}$ site and that connecting the $i\textsuperscript{th}$ site to its nearest neighbor. The radial functions $V_{|m|} (r)$ are given by 
\begin{subequations}
\begin{align}
V_0(r) &= \lambda_0 e^{- \xi_0 \bar{r}^2} \cos(\kappa_0 \bar{r}) \, , \\
V_3(r) &= \lambda_3 \bar{r}^2 e^{- \xi_3 (\bar{r} - x_3)^2} \, ,  \\
V_6(r) &= \lambda_6 e^{- \xi_6 (\bar{r} - x_6)^2} \sin(\kappa_6 \bar{r}) \, .
\end{align}
\label{eqVm}
\end{subequations}
Here $\bar{r} = r/a$. All the parameters appearing above, collectively denoted by $\pi_i(\delta_d)$ where $i = 1, 2, \cdots 10$, are fixed~\cite{PabloPress} using density functional methods and are listed in the Table~\ref{tablepara}. Given the pressure range of interest, the functional dependence of $\pi_i(\delta_d)$ with $\delta_d$ is truncated to a quadratic fit
\beq
\pi_i (\delta_d) = c_i^{(0)} - c_i^{(1)} \delta_d + c_i^{(2)} \delta_d^2 \, .
\label{10para}
\eeq

\begin{table}[b]
\begin{center}
\begin{tabular}{ |c |c| c|  c |}
\hline
$i \, (\pi_i)$ & $c_i^{(0)}$ & $c_i^{(1)}$ & $c_i^{(2)}$ \\
\hline
 $1 \, (\lambda_0)$ & $\phantom{-}0.310$ & $-1.882$ & $\phantom{-}7.741$ \\
 $2 \, (\xi_0)$ & $\phantom{-}1.750$ & $-1.618$ & $\phantom{-}1.848$ \\
 $3 \, (\kappa_0)$ & $\phantom{-}1.990$ & $\phantom{-}1.007$ & $\phantom{-}2.427$ \\
$4 \, (\lambda_3)$ & $-0.068$ & $\phantom{-}0.399$ & $-1.739$ \\
$5 \, (\xi_3)$ & $\phantom{-}3.286$ & $-0.914$ & $\phantom{-}12.011$ \\
$6 \, (x_3)$ & $\phantom{-}0.500$ & $\phantom{-}0.322$ & $\phantom{-}0.908$ \\
$7 \, (\lambda_6)$ & $-0.008$ & $\phantom{-}0.046$ & $-0.183$ \\
$8 \, (\xi_6)$ & $\phantom{-}2.272$ & $-0.721$ & $-4.414$ \\
$9 \, (x_6)$ & $\phantom{-}1.217$ & $\phantom{-}0.027$ & $-0.658$ \\
$10 \, (\kappa_6)$ & $\phantom{-}1.562$ & $-0.371$ & $-0.134$ \\
\hline
\end{tabular}
\caption{Compression, $\delta_d$, dependence of the ten-parameters appearing in Eq.~\eqref{eqVm}. The coefficients, $c_i^{(n)}$, appearing in Eq.~\eqref{10para} are listed below (in eV units). The theory for determining these coefficients was developed in Refs.~\cite{FangKaxiras, PabloPress}.}
\label{tablepara}
\end{center}
\end{table}

For numerical accuracy, our band structure computations are based on this full ten-parameter model (see Fig.~\ref{figBS}); however, for simplicity, henceforth we confine our discussion to an effective one-parameter model. In Eq.~\eqref{eqVm} the strongest contributions to the interlayer tunneling come from the hybridization scales $\lambda_i$. The remaining parameters, the length scales associated with the Wannier orbitals, are weakly dependent on pressure. Thus, a simpler effective model could be constructed by renormalizing these three parameters (first row of Table~\ref{tablepara}), where the renormalization essentially takes into account the angular contributions coming from all the other parameters (the remaining nine rows in Table~\ref{tablepara}). Such an effective set of parameters was obtained in~\cite{PabloPress} by tallying the bandwidth of the flat bands from the ten-parameter model and an \textit{ab initio} $k \cdot p$ model
\begin{gather}
t_\perp (\delta_d) = t_\perp^{(0)} - t_\perp^{(1)} \delta_d + t_\perp^{(2)} \delta_d^2 \quad , \quad \nn \\ 
t_\perp^{(0,1,2)} \simeq (0.31, -1.73, 7.12) \; \text{eV} \, .
\label{effective1para}
\end{gather}
Again, the parameters $t_\perp^{(i)}$s above, which marginally differ from those listed in the first row of Table~\ref{tablepara}, can be simply seen as effective leading parameters after incorporating the angular contributions. Note that the above $t_\perp$, thus constructed, is the single input to the tight-binding Hamiltonian of the previous section.  Now, however, it is dependent on pressure.

\begin{figure}[t]
\centering
\includegraphics[width=0.94 \columnwidth]{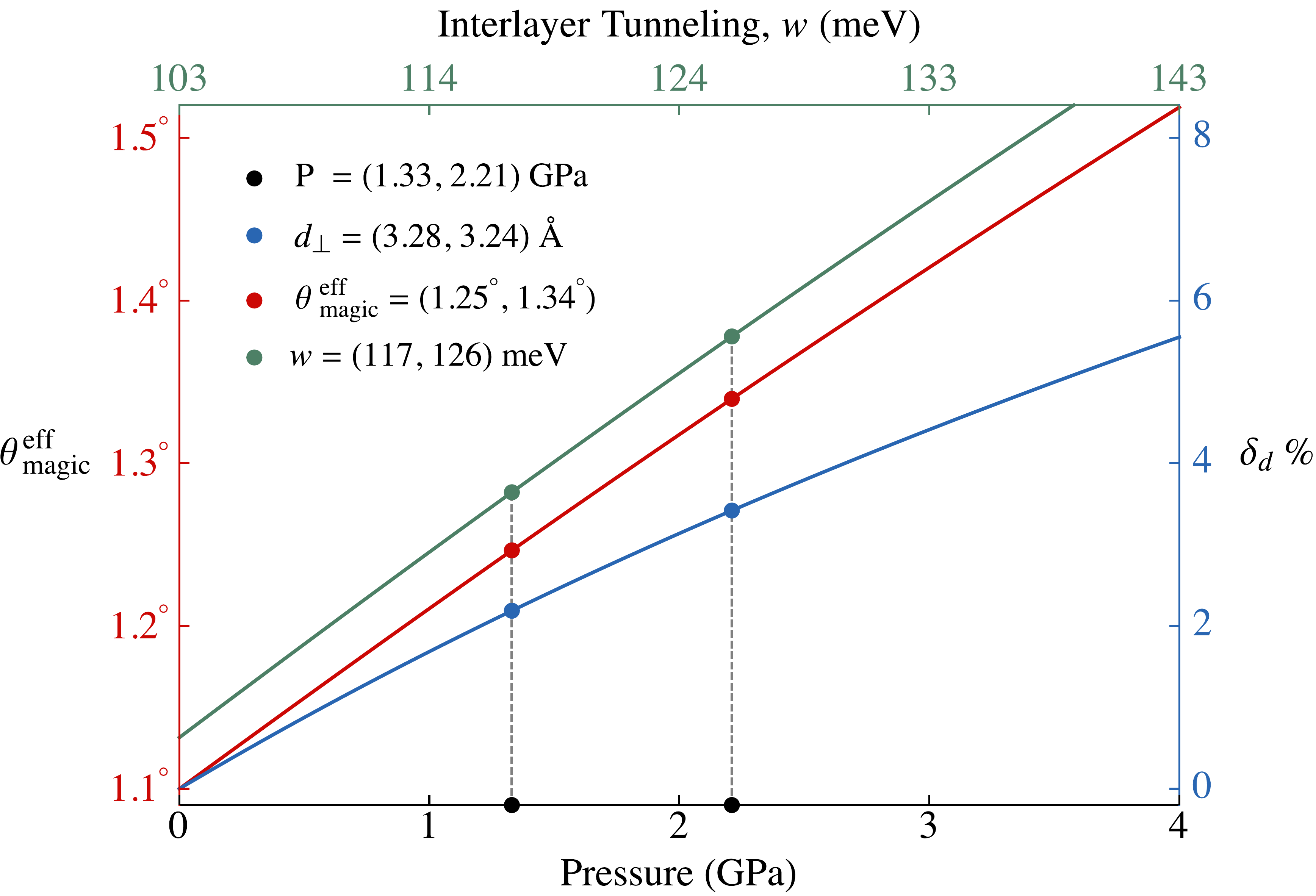} 
\caption{
With increasing external pressure, the inter-layer distance decreases by $\delta_d \%$ (blue curve or right axis) which is described in Eq.~\eqref{Murnaghan}. Reduced separation enhances inter-layer tunneling, $w \approx t_\perp/3$ (green curve or top axis), as can be seen from Eq.~\eqref{effective1para}. This causes an increase (red curve or left axis) in the effective magic angle, $\magic$, where the band become the flattest, see Eq.~\eqref{effectivemagic}. The dots correspond to the reported values of pressure where the measurements of~\cite{YoungDean} were performed. 
}
\label{fig:Press} 
\end{figure}

Using Eq.~\eqref{effective1para}, we now obtain the pressure dependence of $\magic$ discussed before. Note that the magic angle is (roughly) obtained by matching the quasiparticle kinetic energy, $\hbar v_0 K_\theta$, and the hybridization scale, $t_\perp$. Here $v_0 = 10^6\,$m/s is the speed of the electrons in pristine graphene and $K_\theta = 4\pi/3\lambda_s$ is the size of the MBZ. This causes $\magic \sim t_\perp$, or, at ambient pressure, $\theta_{\rm magic} \sim t_\perp^{(0)}$.  Thus, $\hbar v_0 K_\theta = \magic \left(2 t_\perp^{(0)}/ \theta_{\text{magic}} \right) $. Following~\cite{PabloPress, KoreanPress, YoungDean}, we set $\theta_{\text{magic}} = \ang{1.1}$. This gives rise to the following expression for the effective magic angle
\beq
\frac{\magic(P)}{\theta_{\text{\,magic}}} = \frac{t_\perp(P)}{t_\perp^0} =  1 + 5.584 \, \delta_d + 22.97 \, \delta_d^2 \, .
\label{effectivemagic}
\eeq

Figure~\ref{fig:Press} displays the relevant parameters discussed above as functions of external pressure. For a given device with a fixed twist angle $\theta$, which is larger than the ambient pressure magic angle $\theta_{\text{magic}}$, as pressure increases one gradually increases $\magic$. For $\theta = \magic$, one defines the optimum pressure for a particular system, $P_{\rm opt}$ , which is also coincident with the flat-band condition. Increasing the pressure further will relatively tune the system away from the magic angle. The optimal pressure for device D2, for instance, can be solved by demanding $\magic = \ang{1.27}$. From Eq.~\eqref{effectivemagic},  we find that $P \simeq 1.55$ GPa ($\delta_d = 2.5\,\%$). This explains why optimal behavior is seen (among the two available data sets) around $1.33$ GPa ($\delta_d = 2.2\,\%$), as opposed to near $2.21$ GPa ($\delta_d = 3.4\,\%$).

With the use of these parameters, we compute the band structure.   The most notable feature in Fig.~\ref{figBS} is that the bandwidth shrinks as $1.33\,$GPa is approached and increases beyond this pressure.  It is this feature that gives rise to the dome-like shape of the phase diagram of $r_s$ versus hydrostatic pressure, thereby affecting the observed insulating behavior.  Note that, although here we used the tight-binding description, one may also use the effective low energy descriptions developed for ambient pressure~\cite{MCDMoire, LiangMIT, KoshinoFu, AsvinSenthil, Oskar, FaithfulAshvin}; albeit the (tight-binding or continuum) parameters must be fixed taking finite pressure into account.

\section{Computation of  $r_s$}
\label{sec:rs}

\begin{figure}[t]
\centering
\includegraphics[width=0.8\columnwidth]{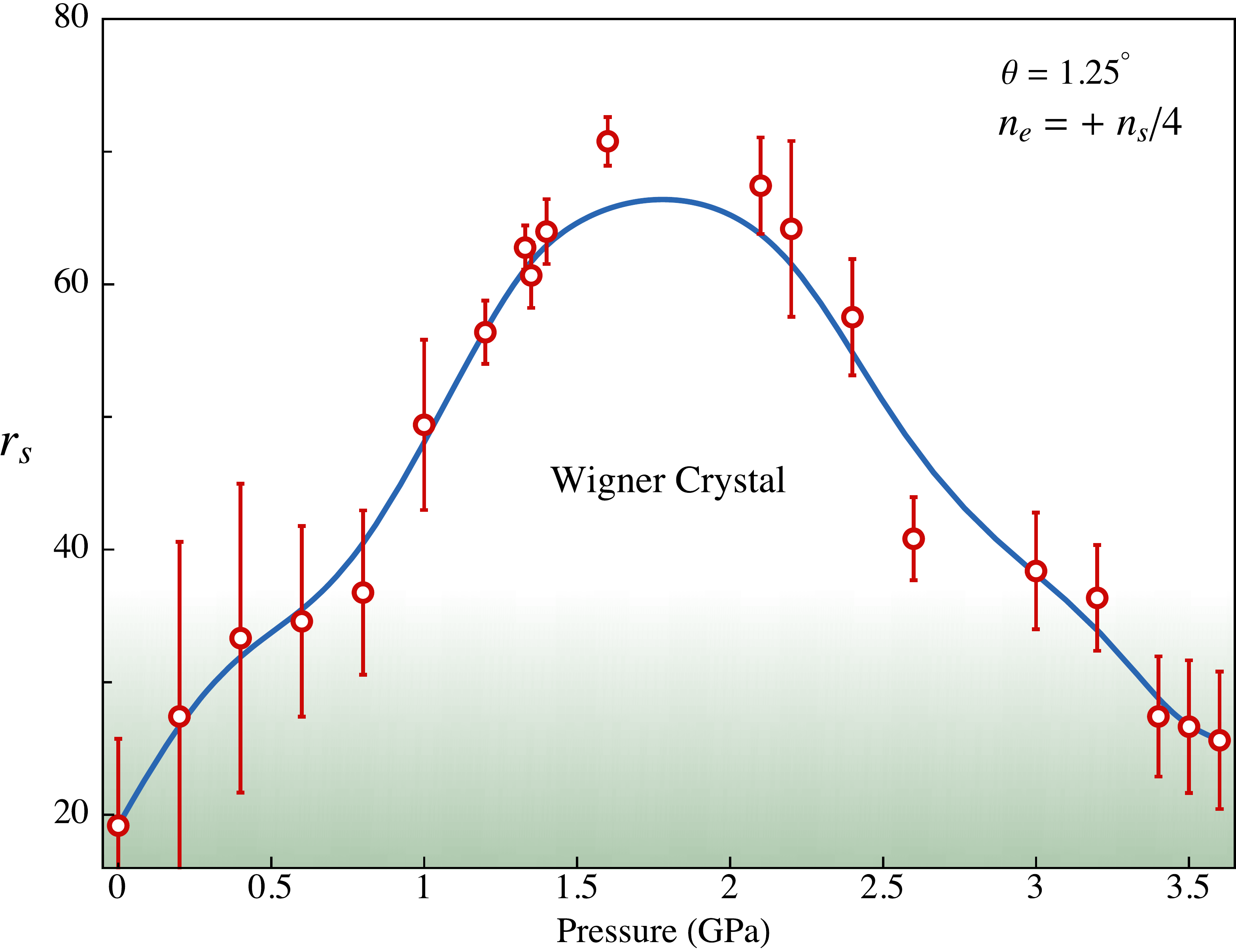} 
\caption{
For the device D2 of~\cite{YoungDean}, we compute $r_s$ (red dots along with computational error associated with coarse graining of the $k$ space) of the quarter-filling state. The blue curve provides a guide for the eye. Its dome-like behavior can explain a similar feature seen in the conductance of the quarter-filling state in the experiment of~\cite{YoungDean}. For the pressure window of $\sim 1-3$ GPa the system enters the Wigner crystallization regime, $r_s \gtrsim 37$. Similar behavior is obtained for $n_e = n_s/2$ and $3n_s/4$.}
\label{fig:rsP} 
\end{figure}

\begin{figure}[t]
\centering
\subfloat[]{\includegraphics[width=0.48 \columnwidth,height=0.46 \columnwidth]{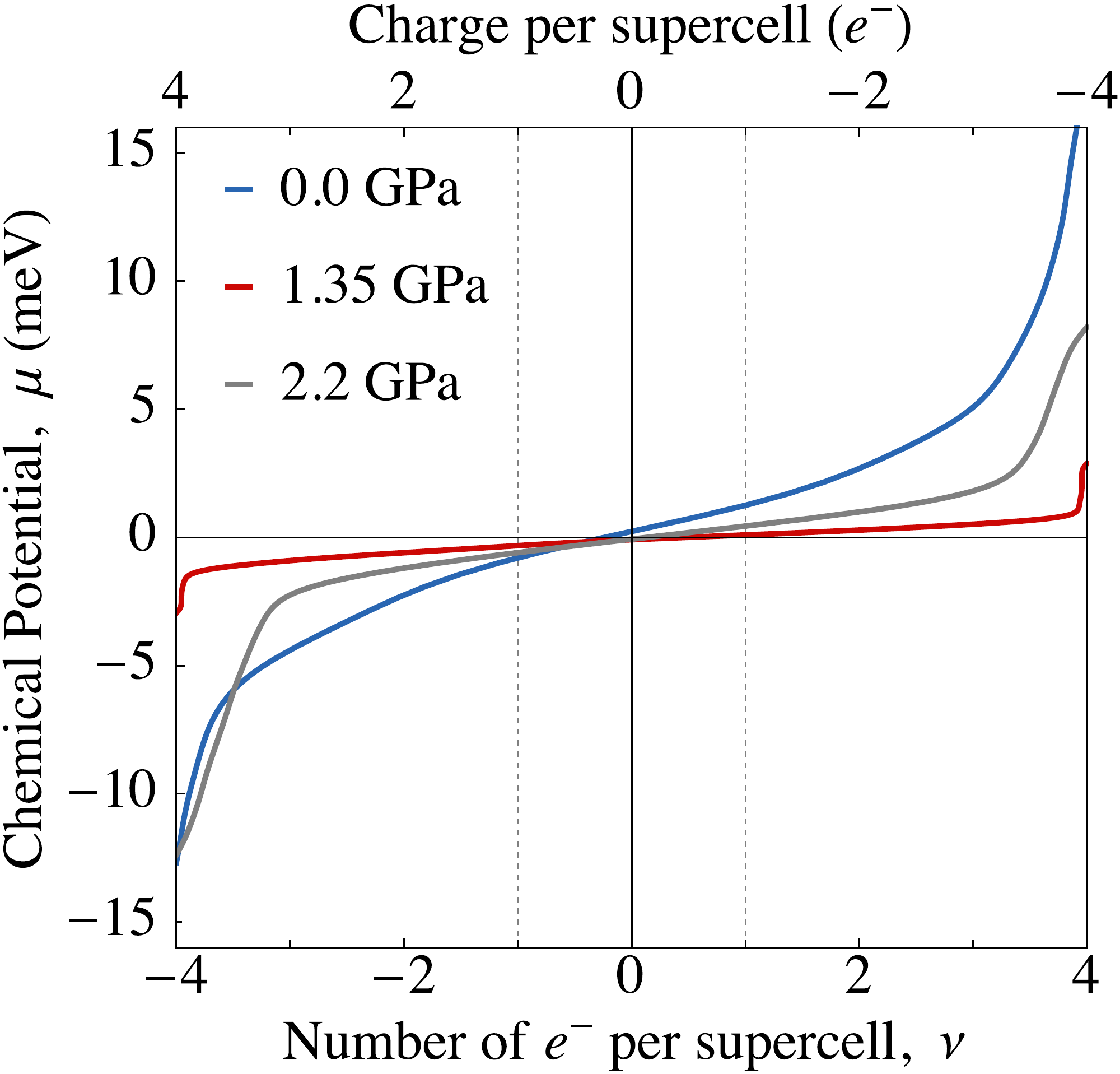} 
\label{cardens}}
\subfloat[]{\includegraphics[width=0.48 \columnwidth,height=0.42 \columnwidth]{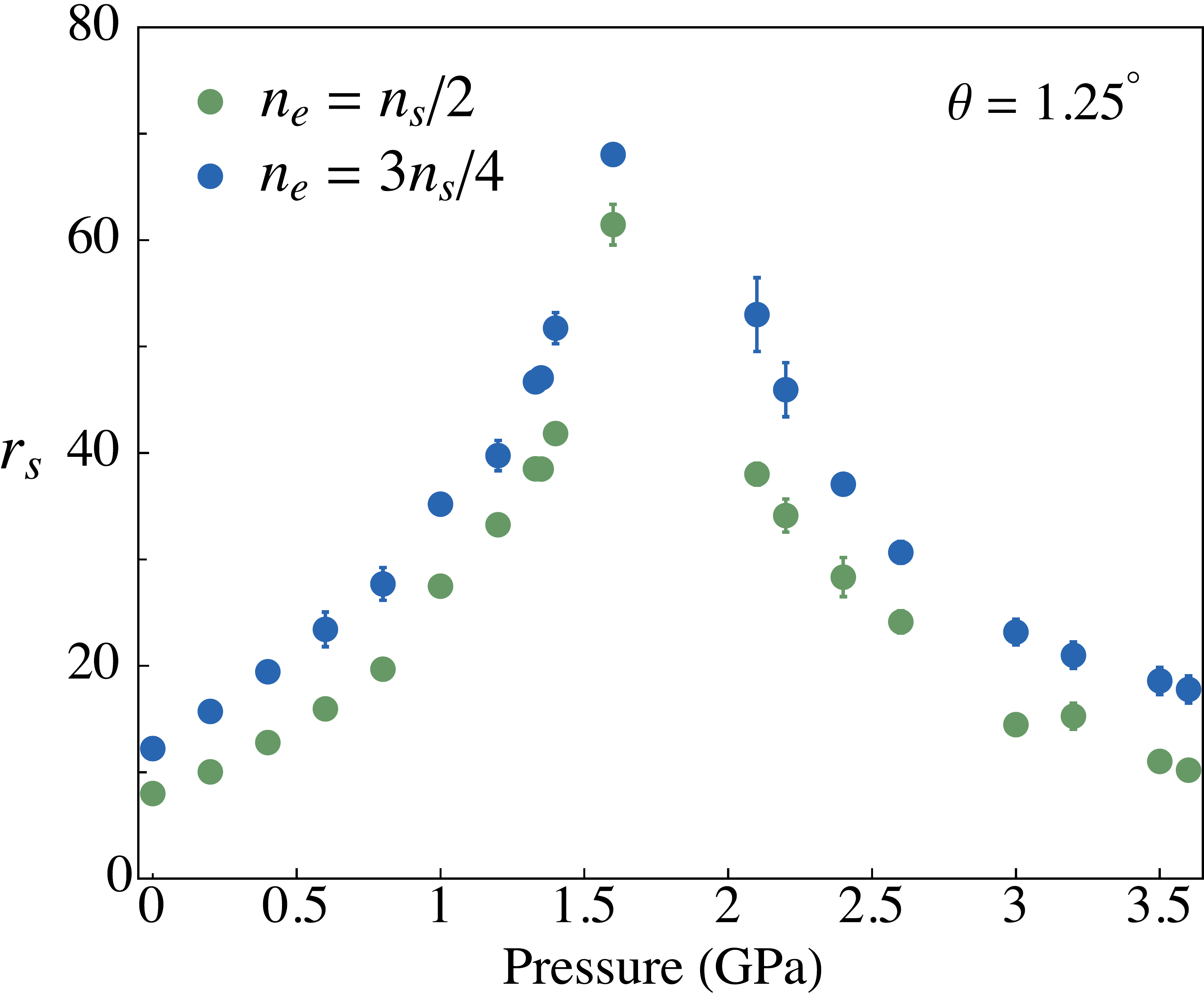} 
\label{rs23}}
\caption{(a) The chemical potential dependence (smoothly fitted) of the carrier concentration for various pressures, as obtained in Eq.~\eqref{nemu}. (b) $r_s$ as a function of pressure obtained for $\nu = 2,3$ states. Similar to the $\nu =1$ state discussed in the main text they also exhibit dome--like behavior.}
\end{figure}

We now turn to the computation of $r_s$.  First we need to estimate Coulomb energy for a TBLG system at $\theta$ angle, 
\begin{align}
E_U = \frac{e^2}{\epsilon r_e} = \left( \frac{\alpha}{\epsilon} \right) \left( \frac{\hbar v_0}{a} \right) \left( \frac{a}{r_e} \right)\, .
\end{align}
Since $v_0$ is about $300$ times smaller than the speed of light, the effective fine structure constant of (suspended) graphene is $\alpha \approx 2.2$. Also note that $\hbar v_0/ a = 2.135\;$eV. In the presence of the hBN substrate, this is reduced by a factor of the effective dielectric constant, $\epsilon = 10$~\footnote
{The dielectric constant of TBLG is largely determined by the encapsulating hBN layers with $\epsilon \sim 6$. Taking screening from the higher bands into account, in the supplementary section of our earlier paper~\cite{ourNano}, we estimated the renormalized $\epsilon$ within the random phase approximation and obtained $\epsilon \sim 10$ near the magic angle. However, since this scheme breaks down in the Wigner regime, the most reliable method to estimate the Coulomb interaction is to use an enhanced dielectric constant, as is customary in the experimental works. A recent discussion on the issue of screening can be found in Ref.~\cite{ScreeningRonny}.}.
The average inter-particle distance can be obtained from $\pi r_e^2 n_e = 1$. For a given filling fraction, $r_e = \sqrt{A_s /\pi \nu} \approx  {0.525} \,\lambda_s/ \sqrt{\nu} $. Combining all of  these expressions, we find that $E_U \approx (15\; \text{meV})  \,\theta^{\,\circ}$. In Sec.~\ref{sec:Correction} we discuss some subtleties involved with a more realistic estimation of $E_U$ in TBLG. The final expression for $r_s$ is 
\begin{align}
r_s \approx  15 \,\text{meV} \, \frac{\theta^{\,\circ}}{E_K} \, \sqrt{\nu} \xrightarrow[\text{D2}]{\text{Device}} \, \frac{20 \,\text{meV}}{E_K \,(\text{meV})} \, \sqrt{\nu} \, .
\label{eq:rs}
\end{align}
In order to fix the kinetic energy above, we first relate the carrier concentration to chemical potential, $\mu$, and since $E_K \lesssim \mu$, for a minimal (and hence conservative) estimate of $r_s$, one can substitute $E_K$ with $\mu$. In order to do so we start by computing the density of states (DOS), $\rho(\epsilon)$, which can be normalized in the following way. Since each moir\'e supercell contains eight electrons at the most, integrating the DOS for the bottom four bands must yield $8$
\beq
\int_{\Lambda_h}^{\Lambda_e} \rho(\epsilon) d\epsilon = 8 \, .
\eeq
Here, $\Lambda_{e, h} \sim \pm 10\, \rm meV$, respectively, provide the upper and lower cutoff for the bottom four bands. Integrating the normalized DOS up to the chemical potential provides the carrier concentration (in order to compare our results with those of ~\cite{YoungDean} we do so for the hole side):
\beq
n_e(\mu)  = \int_{\Lambda_h}^{\mu} \rho(\epsilon) d\epsilon \, .
\label{nemu}
\eeq
This is shown in Fig.~\ref{cardens}.  In obtaining $r_s$ for the $\nu/4$ state one can fix $n_e(\mu) = \nu \, n_s/4$ (e.g., see the gray line for $\nu =1$) and obtain how $\mu$ evolves with pressure along that line. Note the source of error here is the coarse graining of the energy integral above.

In Fig.~\ref{fig:rsP} [or Fig.~\ref{rs23}] we plot the behavior of $r_s$ as a function of pressure, which is clearly dome-like. The key aspect of this figure is the crossing of the Wigner threshold for pressures in the range $0.75 <P<3 $ GPa.   The existence of this window for optimal insulating behavior of the $\nu=1$ state can be tested experimentally.  We find that $P_{opt}=1.5$ GPa, which is close to the experimentally observed optimal pressure, $1.33$ GPa.  Clearly further experiments are needed to map out the non-monotonic dependence of the metal-insulator transition as a function of the uniaxial pressure.  As can be seen in Fig.~\ref{rs23}, similar behavior is seen for the $\nu = 2,3$ states, which, as we showed previously~\cite{ourNano}, correspond to honeycomb and kagome Wigner crystals.

\section{Possible Sources of Corrections to $r_s$}
\label{sec:Correction}

There are several variables present in a realistic hBN-TBLG system which might affect the exactness of our estimated $r_s$. In this section we list numerous such effects and discuss their consequences. We argue that these variables do not influence the order of our estimations significantly nor do they alter our qualitative conclusions. Hence, for simplicity, we have not considered them in our estimation.

Before turning to a discussion pertaining to TBLG we first note that the critical value for Wigner crystallization used in this work, $r^{\rm crit}_s = 37$, though universal across all materials, is somewhat approximate. The computational sources of error are the finite-size effect (extrapolation of finite number of electrons to the thermodynamic limit), and the fitting errors in obtaining $E_0(r_s)$. Cumulatively, they amount to an error of magnitude $\delta r^{\rm crit}_s = 5$ in~\cite{David37}. Another important source of uncertainty in $r_s^{\rm crit}$ is a methodical error arising from the so-called fixed-node approximation used in the diffusion Monte Carlo method~\cite{fixed-node76} (that the actual and the trial wave functions share the same nodal surface). An improved use of this approximation was done in~\cite{DrummondNeeds} by including Slater-Jastrow-backflow wave functions~\cite{Backflow06}. This results in $r_s^{\rm crit} = 31 \pm 1$, which incidentally lowers (raises) the critical pressure at which the metal to insulator (insulator to metal) transition occurs; see Fig.~\ref{fig:rsP}. However, it should also be noted that the transition in~\cite{DrummondNeeds} is from a paramagnetic fluid to a triangular antiferromagnetic crystal, as opposed to a transition from a ferromagnetic fluid to the triangular ferromagnetic crystal as reported in~\cite{David37}.  In view of all these uncertainties the phase boundary in Fig.~\ref{fig:rsP} remains intact.

In our work we have ignored any effects related to atomic relaxation in TBLG. For instance, the optimal lattice configuration of TBLG with twist angle $\lesssim 2^\circ$ is corrugated along the $c$ axis~\cite{Corrugation1}. This causes the inter-layer separation $d_\perp$ (or the inter-layer coupling $w$) to increase (reduce) in the AA-stacking region and decrease (enhance) in the AB/BA-stacking region. The consequences of such an effect on the band structure of near-magic-angle TBLG is that~\cite{KoshinoFu, PabloPress} corrugation significantly enhances the band gap between the moire\'e flat bands and the higher-energy bands, albeit leaving the bandwidth virtually unchanged. Thus, as the bandwidth sets the scale for the kinetic energy as an input into the computation of $r_s$, the effect of out-of-plane relaxation is negligible in our case. Enhancement of the band gap simply strengthens the assumption of the flat bands being isolated.

In-plane relaxation effects  often shrink the area of the AA-stacking region, concomitantly facilitating formation of a triangular domain structure with alternating AB- and BA-stacking regions~\cite{NamKoshino17}. In this case as well, the band gap increases; however, unlike the earlier case, in-plane relaxations cause the bandwidth to increase, though no more than $10 \%$ at ambient pressure. Naively this should also lower our estimations of $r_s (P=0)$ by a similar fraction. However, to the best of our knowledge, the full inclusion of all the relaxation effects (see~\cite{KaxirasBothRelax}), let alone with pressure dependence, has not yet been studied in detail. Thus, for simplicity we ignore any such effects in this work, which can at most change our estimates by 10$\%$.

It must also be noted that most of the near-magic-angle devices suffer from a twist angle inhomogeneity~\cite{PasupathySTM, TwistDis14, TwistDisKim18} which often has dramatic consequences on the phase diagram of TBLG~\cite{EfetovSC}. In other words, the local modulation in the twist angle could render $r_s$ to be a position-dependent function. Thus, it is perfectly possible that the sample as a whole may not undergo crystallization transition but it could form puddles of WCs, phase separated with other insulating or metallic states. Such consideration often plays a key role in experimental observation of WCs~\cite{CompressSLG}.

In all of our calculations, the presence of the hBN layer(s) is accounted for only through the dielectric constant. However, the alignment or misalignment of the hBN substrate with the adjacent graphene layer of TBLG could significantly influence the phase diagram. Most importantly, the appearance or enhancement of a band gap near the Dirac point could~\cite{hBN/G13, hBN/G15} primarily emerge from moir\'e patterns or strains in the bilayer formed out of hBN/graphene~\cite{hBN/G17}. Clearly, such an effect mainly drives the physics near charge neutrality.  For instance, the appearance of a superconducting dome near charge neutrality in~\cite{EfetovSC} could possibly be attributed to the physics of hBN/graphene bilayer. Thus, for the bulk of our interest such an effect does not contribute to $r_s$.  
\section {Concluding Remarks}
\label{sec:Discuss}

We have shown that the pressure dependence of the metal-insulator transition has a natural explanation within the hierarchy of Wigner crystals proposed recently for TBLG~\cite{ourNano}.  
Should the dome-like phase diagram for the $\nu=1$ state be confirmed experimentally, then this would add significant substantiation to the claim that TBLG offers a playground for observing WCs and the possible onset of superconductivity.  

Our proposal that superconductivity lurks in the vicinity of Wigner crystallization is rooted in the retardation effects that are inherent to the strongly correlated regime.  From the potential of interaction of an electron in a Wigner crystal~\cite{Takada},
\beq
V({\bf r})=-\frac32\frac{e^2}{a^\ast r_s}+\frac12\frac{e^2}{(a^\ast r_s)^3}{\bf r}^2=-\frac32\frac{e^2}{a^\ast r_s}+\frac12\omega^2{\bf r}^2,
\eeq
increasing the electron density decreases the restoring frequency, $\omega$, thereby leading to a melting of a WC.  However, when a charge moves in a WC, it must dissociate from the Coulomb or correlation hole that led to the formation of the crystal in the first place.  The size of this correlation hole is $r_e$ and hence is roughly 10, 000 carbon atoms in TBLG at the relevant magic angles.  Such a correlation hole and the electrons move on different time scales.  Once the crystal moves, the correlation hole left behind is now positively charged and hence, on the timescale that it is vacated,  it is attractive to the electrons in its vicinity.  Consequently, such charge retardation effects could mediate pairing.  This is the purely electron analog of the polaron effect and has been proposed previously to mediate pairing in the vicinity of the melting transition of Wigner crystals~\cite{phillipsNature,Takada, BianconiWC}.  Of course the form of the kinetic energy term will have to be modified for TBLG but the content of the argument remains intact.  We hope to address this issue in greater detail in future work.

Regarding the spin dependence of the insulating states, the ferromagnetic triangular WC is well known~\cite{David37} to be energetically favored for the $\nu=1$ state.  The honeycomb WC we proposed has explicitly two electrons residing in each moir\'e cell and hence has $S=0$.  The spin structure of the kagome lattice has no natural singlet correlations and hence should be spin-polarized just as in the $\nu=1$ case. Hence, we anticipate for $\nu=3$ the ground state is a ferromagnet, as has been observed recently~\cite{Kastner}. Previously, ferromagnetic Wigner crystallization has been used to explain the $1/6$-th filling-state in graphyne~\cite{CongjunWu2}. Within a Mott scenario, it is difficult to explain the spin dependence without at the same time invoking sites for the spins which would make the resultant electron lattice distinct from the underlying triangular moir\'e lattice.  Recall, a Mott insulator cannot break any underlying symmetries.  In this regard, the $1/2$-filled honeycomb structures proffered~\cite{Subir,LeeKekule} to explain the $\nu=1/2$ states are instances of the WC we have proposed here. Consequently,  all the features of the novel insulating states in TBLG are captured by a transition to WC.  

It must also be noted that in our proposal, unlike the case of GaAs/AlGaAs heterostructures~\cite{KravchenkoReport} or that of liquid helium~\cite{GrimesAdams}, there is a WC pinned to the underlying moir\'e lattice. This poses a unique set of experimental challenges in distinguishing it from a Mott (or any other correlated) insulator~\cite{DobroMITbook, GabiWCMott_Nature}. For instance, formation of a WC, particularly adjacent to a gapped state, is often signaled by an instability in the thermodynamic compressibility~\cite{EisensteinPRL92}.  Thus far, similar measurements in TBLG~\cite{Spencer} observe phases with nondiverging and non-negative compressibility at commensurate fillings.  Although one cannot rule out the influence of (twist or charge) disorder, strong pinning of the Wigner lattice to the moir\'e lattice may also render the insulating states incompressible~\cite{Future}. Consequently, the precise conclusion to be drawn from the compressibility experiments remains unclear at present.
\section*{acknowledgments}
BP is thankful to Yubo Yang for his generous help with the numerics and for explaining Refs.~\cite{fixed-node76, DrummondNeeds, Backflow06}. We are thankful to Spencer Tomarken and Ray Ashoori for pointing us to Ref.~\cite{CompressSLG} and for discussions regarding Ref.~\cite{Spencer}. We are also thankful to Chandan Setty for his characteristically level-headed remarks and the NSF DMR-1461952 for partial funding of this project.
\\
%
%

%

\end{document}